\documentclass[12pt]{article}
\usepackage{a4}
\usepackage{epsfig,braket,url}
\usepackage{lineno}

\newcommand\diag{\mathop{\mathrm{diag}}}
\newcommand\Tr{\mathop{\mathrm{Trace}}}
\newcommand\hP{\hat P}
\newcommand\hB{\hat B}
\newcommand\hb{\hat b}
\newcommand\tB{\tilde B}
\newcommand\bB{\bar B}
\newcommand\EE{\mathbf{E}}
\begin{document}
\title{Operational improvements for an algorithm to noninvasively
  measure the orbit response matrix in storage rings}
\author{Volker Ziemann, Uppsala University, 75120 Uppsala, Sweden}
\date{\today}
\maketitle
\begin{abstract}\noindent
  We improve the algorithm to noninvasively update the response matrix using
  information from the orbit-feedback system, described in~\cite{ZZ1}. The
  new version is capable of adapting to slow changes of the lattice, albeit
  at the expense of limiting the accuracy. 
\end{abstract}
%
%
\section{Introduction}
\label{sec:intro}
The orbit-response matrix relates the changes in the excitation of steering
magnets to observed position changes on the beam position monitor
system~\cite{MIZI,HUANG}. It is the workhorse needed to correct the beam positions
and to analyze discrepancies between an idealized model of the accelerator to
the ``real'' one using codes like LOCO~\cite{CORBETT,SAFRANEK}. Usually, the
response matrix is either derived from a computer model or it is measured, which
normally requires some dedicated beam time. In~\cite{ZZ1} we presented a method
to improve the response matrix by exploiting correlations between the position
changes and the excitations of steering magnets caused by an orbit-feedback
system. The system runs quasi ``on the side'' and does not perturb the running
accelerator.
\par
Unfortunately the rate of convergence of this system, expecially with very
accurate position monitors, is very slow. Moreover, in~\cite{ZZ1} we assumed that
the model is stationary, which real accelerators, however, often are not;
for example, correcting the tunes or closing the gap of an undulator in
synchrotron light sources slightly affects the beam optics and thereby the
response matrix of the accelerator. To account for these effects, we describe
a modification of the algorithm from~\cite{ZZ1} to make it much more agile to
respond to slow and small changes of the underlying system. We base our discussion
on well-known methods from the theory of system identification described
in~\cite{SYSINF}.
\par
In the next section we briefly review the model and the algorithm as well
as the improvements, before we simulate its performance in the next one.
The tradeoff between speed and accuracy of the algorithm are explored in
Section~\ref{sec:conv} before we come to the conclusions.
\section{Model}
\label{sec:model}
As in~\cite{ZZ1}, we model the dependence of readings from $n$ position
monitors by the $\ket{x}$ on $m$ steering magnet excitations by a dynamical
system
\begin{equation}\label{eq:sys}
  \ket{x_{t+1}}=\ket{x_{t}}+ B \ket{u_t} +\ket{w_t}
  \qquad\mathrm{with}\qquad \ket{u_t}=-K\ket{x_t}\ ,
\end{equation}
where the subscript $t$ denotes a discrete time step from one iteration to
the next, $B$ is the $n\times m$ dimensional orbit response matrix, $K$ is
the $m\times n$-dimensional correction matrix of the orbit correction system,
and $\ket{w_t}$ describes noise in the system, characterized by the expectation
value $\mathbf{E}\{\ket{w_s}\bra{w_t}\}=\sigma_w^2\delta_{st}\mathbf{1}$. Here
$\mathbf{1}$ is the $n\times n$ unit matrix, $\delta_{st}$ is the Kronecker
symbol, and $\sigma_w$ is the rms magnitude of the noise.
We borrow the notation with
bra and ket vectors from quantum mechanics, because keeping track of many
inner and outer products becomes transparent. Here a ket denotes a column
vector and a bra denotes a row vector. Throughout this report, the notation
is consistent with~\cite{ZZ1}.
\par
Our task is now to determine an estimate $\hB^{ij}_T$ of the matrix elements
$B^{ij}$ from recordings of all monitor readings $x_{t}^i$ with $1\leq i\leq n$
and steerer excitations $u_t^j$ with $1\leq j\leq  m$. Here the subscripts
denote times steps and superscripts label monitors and steerers. We point out
that the estimated matrix $\hB$ depends on the time step $T$ and typically
improves as more samples are included when $T$ grows. Note the caret to
indicate that $\hB$ is an estimate.
\par
To this end we employ standard methods from the theory of system
identification~\cite{SYSINF,LJUNG} and write Equation~\ref{eq:sys} for one
monitor labeled $i$
\begin{equation}
  x^i_{s+1} - x^i_s
  =\left(
    \begin{array}{ccc}
           u^1_s & \dots & u^m_s \\
    \end{array}
  \right)
  \left(\begin{array}{c} \hB^{i1} \\ \vdots \\ \hB^{im}\end{array}\right)\ ,              
\end{equation}
which provides us with information about row $i$ of $\hB$. Stacking many copies
of this equation for successive time steps $1\leq s \leq T$ on top of each other
leads to
\begin{equation}\label{eq:defU}
  \left(\begin{array}{c} x^i_{2} - x^i_1\\ \vdots \\ x^i_{T+1} - x^i_T\end{array}\right)         
  = U_T
  \left(\begin{array}{c} \hB^{i1}_T \\ \vdots \\ \hB^{im}_T\end{array}\right)
  \qquad\mathrm{with}\qquad
  U_T  =\left(
    \begin{array}{ccc}
           u^1_1 & \dots & u^m_1 \\
           & \vdots & \\
           u^1_T & \dots & u^m_T \\      
    \end{array}
  \right)\ .
\end{equation}
As $T$ increases the matrix $U_T$ grows by one line in each time step and we gather
more and more information about the row $i$ of $\hB_T$ after time step $T$. In this
way Equation~\ref{eq:defU} becomes a highly overdetermined linear system that can
be solved in the least-squares sense by the pseudo inverse~\cite{FYF}
\begin{equation}\label{eq:defB}
  \left(\begin{array}{c} \hB^{i1}_T \\ \vdots \\ \hB^{im}_T\end{array}\right)
  =\left(U_T^{\top}U_T\right)^{-1}U_T^{\top}
  \left(\begin{array}{c} x^i_{2} - x^i_1\\ \vdots \\ x^i_{T+1} - x^i_T\end{array}\right)\ .      
\end{equation}
Of course we have to repeat the same procedure for all other rows of $\hB_T$ to
obtain the complete estimate of the response matrix after $T$ time steps. 
Equation~\ref{eq:defB} describes a linear map from the vector with the position
differences on the right-hand side onto the vector with row $i$ of $\hB_T$.
Therefore~\cite{FYF} $P_T=\left(U_T^{\top}U_T\right)^{-1}$ is the empirical (data-driven)
covariance matrix of the $\hB_T$ after multiplying with the error bars of the
positions, which is $\sigma^2_w$. The error bars $\sigma(\hB)$ of the fitted
$\hB_T$ are therefore approximately given by the square root of the diagonal
elements of $\sigma_w^2\left(U_T^{\top}U_T\right)^{-1}$ up to a factor of order
unity.
\par
Instead of storing and inverting $U_T$ after each times step, we employ the
Sherman-Morrison formula~\cite{SHEMO} to iteratively update $\hB_T$ and the
empirical covariance matrix $P_T=\left(U_T^{\top}U_T\right)^{-1}$ that appears in
Equation~\ref{eq:defB}. In each time step the row vector $\bra{u_{T+1}}=(u^1_{T+1},
\dots,  u^m_{T+1})$ is added to the bottom of $U_T$ which allows us to write
$P_{T+1}^{-1}=P_T^{-1}+\ket{u_{T+1}}\bra{u_{T+1}}$.
\par
In contrast to~\cite{ZZ1}, here we introduce a factor $\alpha=1-1/N_f$ that
weighs down the older samples~\cite{SYSINF2}, where $N_f$ is the exponential
time constant (in units of iterations) that controls this ``forgetting.'' We
therefore write
\begin{equation}\label{eq:PTa}
  P_{T+1}^{-1}=\alpha P_T^{-1}+\ket{u_{T+1}}\bra{u_{T+1}}
\end{equation}
whereas in~\cite{ZZ1} we had $\alpha=1.$ Note that $0\leq \alpha\leq 1$ assigns
a weight to all rows of $U_T$, except the most recent one.
\par
All derivations from~\cite{ZZ1} to invert Equation~\ref{eq:PTa} are still valid,
provided we substitute $P_T \to P_T/\alpha$ in Equations~6 and~7 from~\cite{ZZ1}.
After some straightforward algebra we obtain for the updated empirical covariance
matrix $P_{T+1}$
\begin{equation}\label{eq:upP}
P_{T+1} = \frac{1}{\alpha}\left[P_T - \frac{P_T\ket{u_{T+1}}\bra{u_{T+1}}P_T}{\alpha+\braket{u_{T+1}|P_T|u_{T+1}}}\right]\ .
\end{equation}
and for the updated response matrix $\hB_{T+1}$
\begin{equation}\label{eq:upB}
  \hB_{T+1}=\hB_T+\frac{\left(\ket{x_{T+2}}-\ket{x_{T+1}}-\hB_T\ket{u_{T+1}}\right)\bra{u_{T+1}}P_T}
  {\alpha+\braket{u_{T+1}|P_T|u_{T+1}}}\ .
\end{equation}
We refer to Appendix~A and~B in~\cite{ZZ1}, as well as~\cite{SYSINF2} for more
details. We point out the enormous advantage of the iterative procedure to update
$P_T$ and $\hB_T$ over repeatedly solving Equation~\ref{eq:defB} for $\hB_{T+1}$.
Here we only have to store $P_T$ and $\hB_T$ in memory and update them with
Equations~\ref{eq:upP} and~\ref{eq:upB} as new information represented by the
monitor readings and corresponding steerer excitations becomes available.
\par
The following code snippet illustrates how to implement the algorithm in MATLAB
\begin{verbatim}
  function [Bhatnew,Pnew,xnew]=one_iteration(Bhat,P,x,alpha)
  global sig Btilde Breal K 
  u=-K*x;                                    % eq. 1, second part
  xnew=x+Breal*u+sig*randn(size(x));         % eq. 1, first part
  tmp=u'*P;                                  % <u|P
  denominv=1/(alpha+tmp*u);                  % 1/(alpha+<u|P|u>)
  Pnew=(P-tmp'*tmp*denominv)/alpha;          % eq. 6
  Bhatnew=Bhat+(xnew-x-Bhat*u)*tmp*denominv; % eq. 7
\end{verbatim}
where all variables are consistently named to those used in the text. In the next
section we explore the algorithm with numerical simulations.
\section{Simulation}
\label{sec:sim}
We test the updated algorithm with the same model used in~\cite{ZZ1}; a FODO ring
with ten cells having phase advances of  $\mu_x/2\pi=0.228$ in the horizontal
plane and $\mu_y/2\pi=0.238$ in the vertical. A position monitor and steerer
are placed at the same location as the (thin quad) focusing quadrupoles. We
use the ``ideal'' response matrix $\tB$ to calculate the correction matrix
$K=\left(\tB^{\top} \tB\right)^{-1}\tB^{\top}$ that appears in Equation~\ref{eq:sys}.
We then randomly vary the focal lengths of all quadrupoles by 5\,\% to create
a ``real'' response matrix $B$. One of the quadrupoles is varied by an
additional 5\,\%, which results in a second ``real'' response matrix $\bB$
that we will use as an example to model changes to the beam optics. In all
simulations we use $\sigma_w=0.1\,$mm to quantify the monitor errors.
\par
\begin{figure}[tb]
  \begin{center}
    \includegraphics[width=0.9\textwidth]{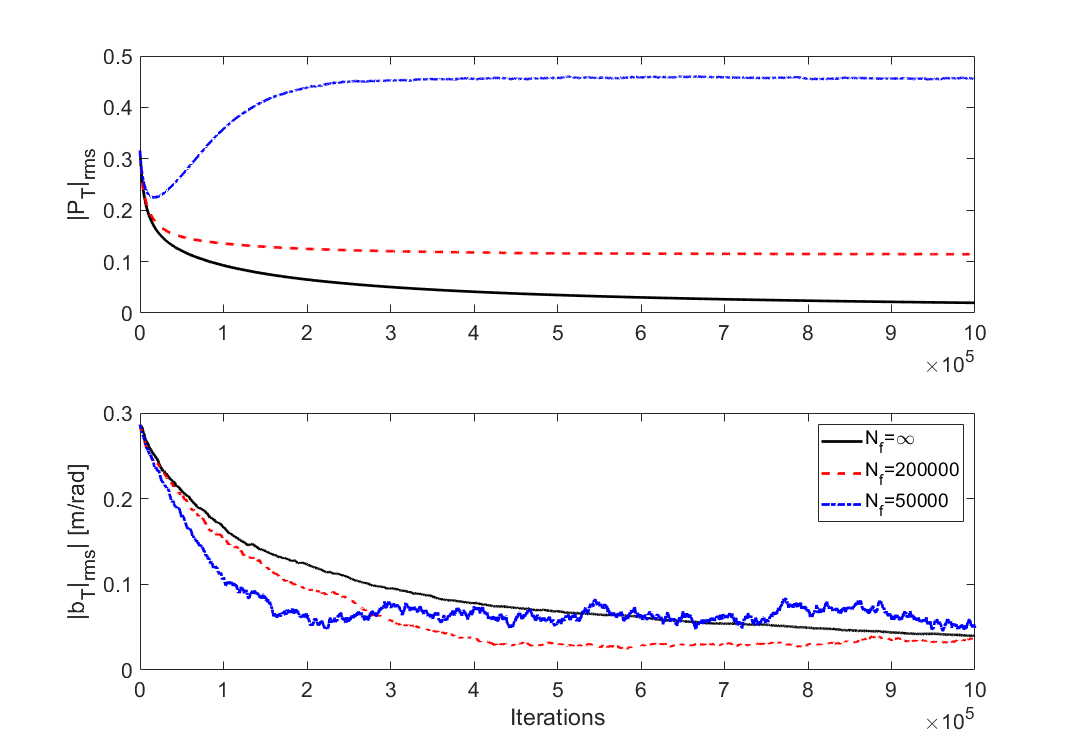}
  \end{center}
  \caption{\label{fig:simc}$|P_T|_{rms}$ (top) and $|b_T|_{rms}$ (bottom) as a
    function of the iterations for $N_f=\infty$ (solid black), $N_f=200\,000$
    (dashed red), and $N_f=50\,000$ (dot dashed blue).}
\end{figure}
In order to assess the performance of our algorithm we introduce the
estimation error $b_T=\hB_T-B$ as the difference between estimate $\hB_T$
and the ``real'' response matrix $B$ (or $\hB$). The rms value of all
its matrix elements $|b_T|_{rms}$ can be calculated from
\begin{equation}
  |b_T|_{rms}=\sqrt{\frac{\Tr\left((\hB_T-B)^{\top}(\hB_T-B)\right)}{nm}}\ .
\end{equation}
In the same fashion, we introduce $|P_T|_{rms}=\sqrt{\Tr\left(P_T^{\top}P_T\right)/m^2}$.
\par
In a first simulation, we initialize $\hB$ with the ``ideal'' matrix from the
computer model, while we use the real matrix $B$ to model the response of the
``real'' system with equation~\ref{eq:sys}. The upper panel in Figure~\ref{fig:simc}
shows $|P_T|_{rms}$ for one million iterations and the lower panel shows $|b_T|_{rms}$.
Each panel shows curves for three values of the forgetting parameter $N_f$. The
black curves correspond to $N_f=\infty$ or $\alpha=1$, the case already covered
in~\cite{ZZ1}. The red dashed curves correspond to $N_f=200\,000$ and the blue dot-dashed
curves to $N_f=50\,000$. From the lower panel we observe that decreasing values of
$N_f$ indeed cause $|b_T|_{rms}$ to decrease more quickly, albeit at the expense of
an deteriorated asymptotic behavior. The read and blue curves no longer approach
zero, as the black one was shown to do in~\cite{ZZ1}. This observation is consistent
with the evolution of $|P_T|_{rms}$ shown in the upper panel. Instead of decreasing
to zero, as the black curve does, the read and blue curves asymptotically approach
finite limiting values. Considering that $P_T$ is the empirical covariance matrix
that describes the error bars of the $\hB_T$ we cannot expect them to approach
$B$ arbitrarily close, as they do with $N_f=\infty$.
\par
\begin{figure}[tb]
  \begin{center}
    \includegraphics[width=0.47\textwidth]{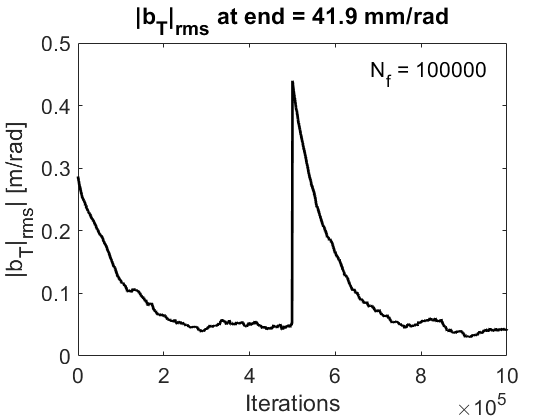}
    \includegraphics[width=0.47\textwidth]{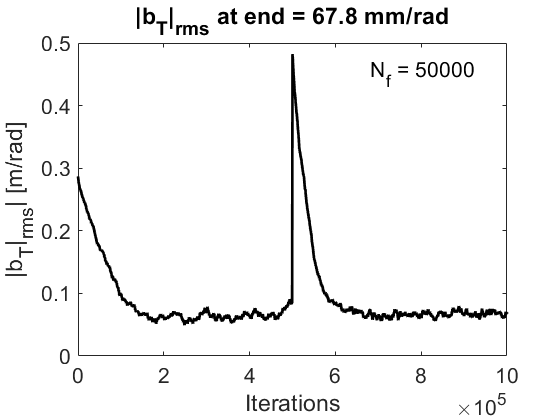}
    \includegraphics[width=0.47\textwidth]{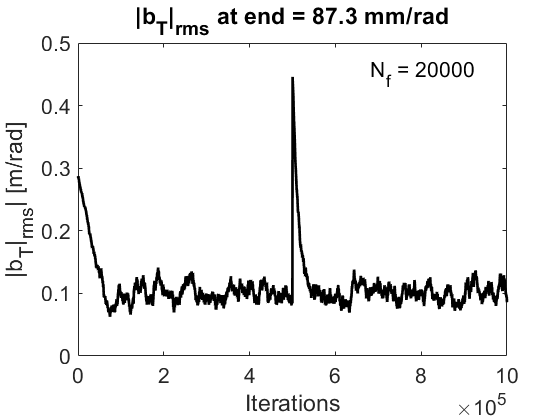}
    \includegraphics[width=0.47\textwidth]{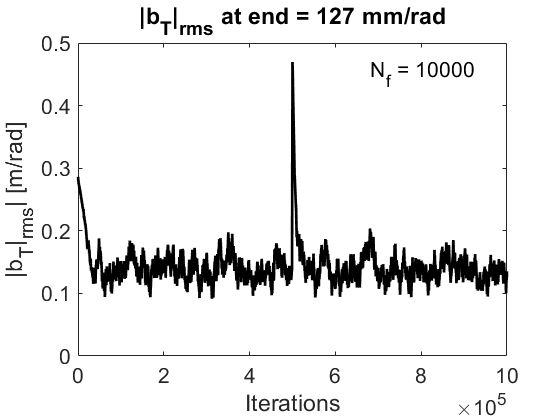}
  \end{center}
  \caption{\label{fig:sim0}$|b_T|_{rms}$ for the four values of $N_f$ indicated
    in the top right corner of the respective plots. After 500\,000 iterations
    the ``real'' system matrix $B$ is replaced by $\bB$ which causes the spike
    in the middle of the plots.}
\end{figure}
Figure~\ref{fig:sim0} shows $|b_T|_{rms}$ for one million iterations for
six values of $N_f$ where, after 500\,000 iterations we replace the ``real''
response matrix $B$ by $\bB$, which is derived from a lattice with one
quadrupole value changed by 5\,\%, as mentioned before. At the same time,
we also replace $B$ by $\bB$ when calculating $|b_T|_{rms}$, because after
the new optics is in place, we expect the algorithm to converge towards
$\bB$, rather than $B$. The values of $N_f$ are indicated on the top left
corner of the six plots. We observe that the algorithm in all cases uses
the first half of the plot to approach $B$, already seen in
Figure~\ref{fig:simc}. But after the new optics is in place after 500\,000
iterations, the best current approximation $\hB_T$ differs significantly
from new reference $\bB$, which causes the spike in the middle part of
all the plots. Forgetting the old configuration and approaching the new
reference happens on the time scale given by $N_f$. Smaller values lead
to a faster approach. Again, at the expense of a faster approach being paid
for by an elevated asymptotic level, which we note in the title bar of
each plot to vary from around 40\,mm/rad for $N_f=100\,000$ to more than
120\,mm/rad for $N_f=10\,000$.
\par
\begin{figure}[tb]
  \begin{center}
    \includegraphics[width=0.8\textwidth]{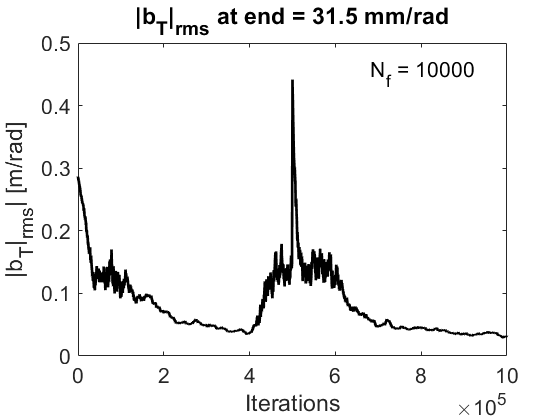}
  \end{center}
  \caption{\label{fig:sim1}$|b_T|_{rms}$ for one million iterations while
    the ``real'' system matrix $B$ is replaced by $\bB$ after $5\times 10^5$
    iterations. For the first $10^5$ iterations we use $N_f=10\,000$, then
    use $N_f=200\,000$ between $10^5$ and $4\times 10^5$ iterations, where
    we again set $N_f=10\,000$ while the change of the system matrix occurs.
    After $6\times 10^5$ iterations we set $N_f=200\,000$.}
\end{figure}
In order to remedy the deteriorated asymptotic level we explored whether it is
possible to temporarily adjust $N_f$ to allow the system to quickly react to
anticipated changes in the lattice, for example, to accommodate an undulator gap
to be closed. Figure~\ref{fig:sim1} shows the configuration leading to the
bottom right plot in Figure~\ref{fig:sim0} with $N_f=10\,000$, only here
temporarily increase $N_f$ to 200\,000 between iteration 100\,000 and 400\,000
and again after iteration 600\,000. We clearly see that the approximation
gets better during the windows with $N_f=200\,000$. Once $N_f$ is decreased
to 10\,000 less information is available and the approximation gets worse, even
before the change that causes the spike. But the system is much more agile to
react and quickly adapts to the new system, albeit with bad precision until
the larger values of $N_f$ after iteration 600\,000 improves the estimate
significantly.
\par
Based on the discussion in this section, we suggest to adapt $N_f$, and thereby
$\alpha$ to the anticipated running mode of the accelerator. If there is
along period  of tranquility, a large value is beneficial, only to be changed
once more activity, for example, tune corrections or changes of undulator gaps
are imminent.
\par
In the next section we will theoretically analyze the time-dependent behavior
of the system.
\section{Convergence}
\label{sec:conv}
\begin{figure}[tb]
  \begin{center}
    \includegraphics[width=0.47\textwidth]{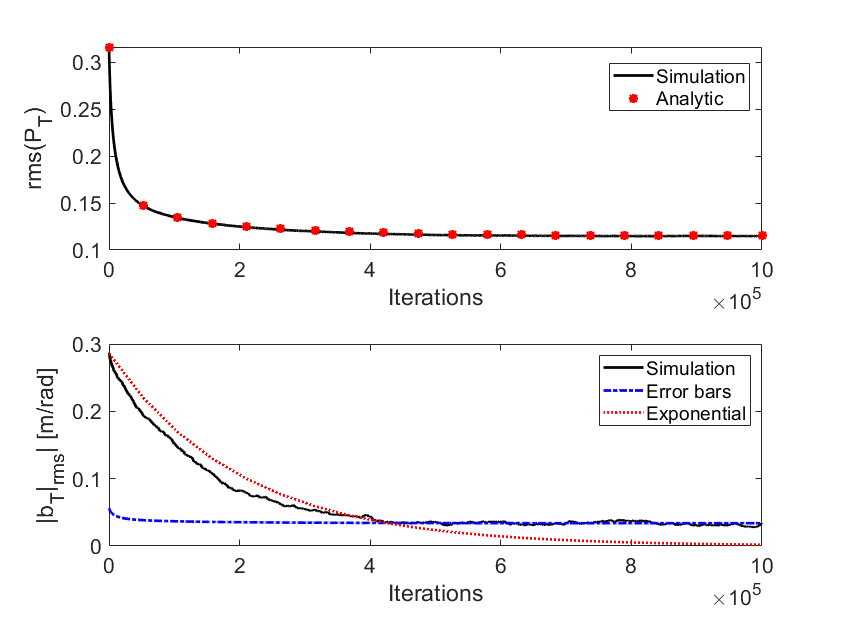}
    \includegraphics[width=0.47\textwidth]{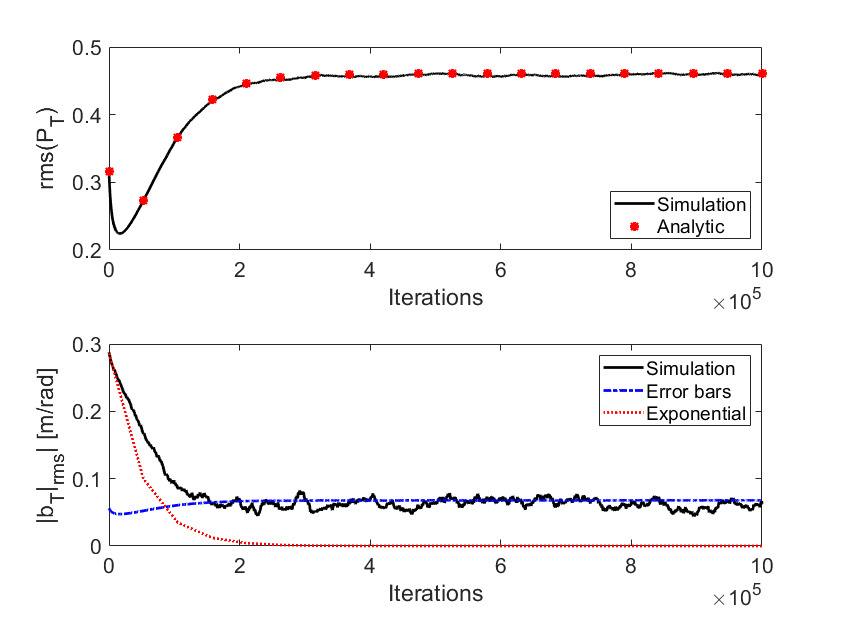}
  \end{center}
  \caption{\label{fig:ptana}Simulations for $|P_T|_{rms}$ (top) and $|b_T|_{rms}$ (bottom)
    as well as analytic results as a function of the iterations for $N_f=2\times 10^5$
    (left) and $N_f=5\times 10^4$ (right).  Note how $|b_T|_{rms}$ initially follows and
    exponential form until it becomes comparable to the noise floor, indicated by the
    blue dot-dashed line.}
\end{figure}
Equations~\ref{eq:upP} and~\ref{eq:upB} describe the time dependence and
thus also the convergence of the response matrix towards the ``real'' one.
We first consider Equation~\ref{eq:upP}, because it only depends on the
most recent steerer excitation through $\ket{u_{T}}\bra{u_{T}}$, where $\ket{u_T}
=-K\ket{x_T}$. Orbit correction systems are always configured such that
the correction matrix $K$ is closely related to the inverse of the
response matrix $B$, such that the largest eigenvalue of $1-BK$ is small.
By virtue of $\ket{x_T}=(1-BK)\ket{x_{T-1}}+\ket{w_{T-1}}$ this implies
that the most recent position $\ket{x_T}$ holds no or very little memory
of all previous position and is dominated by noise $\ket{w_{T-1}}$. As
a consequence we find
$Q=\EE\left\{\ket{u_{T}}\bra{u_{T}}\right\} \approx\sigma_w^2K K^{\top}$ is a
constant matrix. See~\cite{ZZ1} for a more detailed discussion and how to
include small additional variations of the steerers---so-called dithering---in
the analysis.
\par
We now insert this averaged matrix into Equation~\ref{eq:upP} and find
\begin{equation}\label{eq:upPQ}
  \hP_{T+1} = \frac{1}{\alpha}\left[\hP_T - \frac{\hP_TQ\hP_T}{\alpha+\Tr(Q\hP_T)}\right]\ ,
\end{equation}
where we rewrite the expectation value in the denominator as a trace. We placed
a caret over $P_T$ to distinguish it from the solution of Equation~\ref{eq:upP}.
In order to solve this system we now neglect the denominator, which is close to
unity and we are left with $\hP_{T+1}=(\hP_T - \hP_TQ\hP_T)/\alpha$ and by 
subtracting $\hP_T$ on both sides we obtain
$d\hP_T/dT=\hP_{T+1}-\hP_T=(\hP_T -\alpha\hP_t- \hP_TQ\hP_T)/\alpha$. Here we
also introduced the difference between $\hP_T$ in two times steps as a
differential. We observe that the
matrix $Q$ is by construction symmetric and we can therefore diagonalize it,
which leads to $Q=O D O^{\top}$ with $D=\diag(\lambda_1,\dots,\lambda_m)$ and
an orthogonal matrix $O$. The starting matrix $\hP_0$ is the unit matrix and
always diagonal. Therefore Equation~\ref{eq:upPQ} can be written as $m$ independent
equations for each of the diagonal elements $p_{j,T}$ of $\hP_T$. The differential
equation for $\hP_T$ thus defines the corresponding one for each of the modes
with its particular eigenvalue $\lambda_j$
\begin{equation}
  \frac{dp_{j,T}}{dT}=\left(\frac{1-\alpha}{\alpha}\right) p_{j,T}
  - \left(\frac{\lambda_j}{\alpha}\right) p_{j,T}^2\ ,
\end{equation}
which has the solution
\begin{equation}\label{eq:PTT}
  p_{j,T}=\frac{\beta}{\lambda_j/\alpha+(\beta/x_0-\lambda_j/\alpha)e^{-\beta T}}
\end{equation}
with the abbreviation $\beta=(1-\alpha)/\alpha=1/(N_f-1)$. We clearly see that
asymptotically $p_T$ approaches the finite limit $p_{j,\infty}=\alpha\beta/\lambda_j
=1/(N_f\lambda_j)$. Considering that the error bars of reconstructed response
are given in terms of $P_T$ we can expect that increasing $N_f$ improves the
approximation. Note that, apart from $N_f$, only the eigenvalues $\lambda_j$
of $Q=\sigma_w^2K K^{\top}$ enter. In particular, the asymptotic values are
therefore inversely proportional to $\sigma_w$; the algorithm works better with
noisy monitors, because it ``learns from noise.'' From the $p_{j,T}$ we can
reconstruct $\hP_T$ from
\begin{equation}\label{eq:ptanaly}
  \hP_t = O \diag(p_{1,T},\dots, p_{m,T}) O^{\top}
\end{equation}
from which we derive $|\hP_T|_{rms}$ in the same was as for $|P_T|_{rms}$ that comes
from the  numerical simulation. The upper panels in Figure~\ref{fig:ptana} show them for
$N_f=2\times 10^{5}$ on the left and $N_f=5\times 10^{4}$ on the right. The agreement
between simulation and Equation~\ref{eq:ptanaly} in both cases is very good. Also
the approach to finite asymptotic values is clearly visible. 
\par
Substituting $\ket{x_{T+2}}-\ket{x_{T+1}} =B\ket{u_{T+1}}$ in Equation~\ref{eq:upB}
allows us to analyze the convergence of $\hB_T$ towards $B$ from
\begin{equation}\label{eq:dBiter}
  \left(\hB_{T+1} -B\right) = \left(\hB_T-B\right) - \left(\hB_T-B\right)
  \frac{\ket{u_{T+1}}\bra{u_{T+1}}P_T}{\alpha+\braket{u_{T+1}|P_T|u_{T+1}}}\ .
\end{equation}
We now introduce $\hb_T=\hB_T -B$ to simplify writing and omit the denominator with
the trace, as before. Moreover, we replace $P_T$ by its approximation $\hP_T$ and
replace $\ket{u_{T+1}}\bra{u_{T+1}}$ by its expectation value $Q$, which brings us
to $\hb_{T+1}=\hb_T(1-Q\hP_T)$ and by turning the difference equation into a
differential equation with $\hb_{T+1}-\hb_T\approx d\hb_T/dT$ to
$d\hb_T/dT = -\hb_T Q \hP_T$. Despite  $\hb_T$ not being simultaneously diagonal
with $\hP_T$ and $Q$, we make the daring assumption that there are corresponding
modes with eigenvalues $\kappa_{j,T}$, such that we can write 
\begin{equation}
  \frac{d\kappa_{j,T}}{\kappa_{j,T}} = -\lambda_j p_{j,T} dT
  =  -\frac{\lambda_j\beta dT}{\lambda_j/\alpha+(\beta/p_0-\lambda_j/\alpha)e^{-\beta T}}\ ,
\end{equation}
where we substituted $p_{j,T}$ from Equation~\ref{eq:PTT}. Integrating both sides,
where we note that the integral on the right-hand side is elementary, we find
\begin{equation}
  \log\left(\frac{\kappa_{j,T}}{\kappa_{j,0}}\right)
  = -\lambda_j\beta\left[\frac{\alpha T}{\lambda_j} + \frac{\alpha}{\beta\lambda_j}
    \log\left(\frac{\lambda_j}{\alpha}+\left(\frac{\beta}{p_0}-\frac{\lambda_j}{\alpha}\right)e^{-\beta T}\right)\right]
  \approx-\alpha\beta T 
\end{equation}
where we only kept the term linear in $T$ as the leading contribution. Replacing
$\alpha\beta=1/N_f$ we see that the time scale on which the difference between $\hB_T$
and the real response matrix $B$ vanishes is given by $e^{-T/N_f}$, at least in the
dominant order. Since this applies to all modes $\kappa_j$ we feel that the daring
assumption is acceptable.
\par
On the bottom panels in Figure~\ref{fig:ptana} we show the evolution
of $|b_T|_{rms}$ coming from a simulation as black lines and $|b_0|_{rms}e^{-T/N_f}$
as the dot-dashed red line which shows a reasonable agreement. We also observe that
the exponential reduction only works during the initial phase until $|b_T|_{rms}$
becomes comparable to the error bars that are proportional to the magnitude of the
matrix elements of the empirical covariance matrix $|P_T|_{rms}$. We therefore also
show the $\sqrt{|P_T|_{rms}}\sigma_w$ as an indication of these error bars. Once the
exponential part of the convergence comes to a point where $|b_T|_{rms}$ becomes
comparable to  $\sqrt{|P_T|_{rms}}\sigma_w$ it no longer improves. The only way at
this point is to increase $N_f$ to reduce the asymptotic values of $|P_T|_{rms}$
and thus lowers the noise floor, which allows $|b_T|_{rms}$ to decrease further.
But this is just what Figure~\ref{fig:sim1} shows.
\par
In the simulations we used the same number of monitors and steerers ($n=m$),
but this restriction can be overcome using the methods discussed in Section~VII
in~\cite{ZZ1} that we do not repeat here. Other aspects discussed there, such
as delays in the system, remain equally valid. 
\section{Conclusions}
\label{sec:con}
We presented an improved version of the algorithm to noninvasively measure the
orbit response matrix in storage rings. Following~\cite{SYSINF2} we introduce
a time horizon $N_f$ after which the algorithm ``forgets'' old information,
which makes it much more agile to respond to new information, for example, due
to a changed ``real'' response matrix. We found the time constant (in numbers
of iterations) of convergence, at least initially, is given by $N_f$, smaller
values are favorable. On the other hand, we also found that the asymptotically
achievable accuracy is proportional to $1/N_f$, thus favoring large values
of $N_f$.
\par
It is however, possible, to dynamically adjust $N_f$ to the prevailing conditions
of operation. In long periods of tranquility $N_f$ can be increased, only to
reduce it, once changes to the accelerator configuration are expected.
\par
We gratefully acknowledge fruitful discussions with Ingvar Ziemann, University
of Pennsylvania in Philadelphia.
%
%
%
\bibliographystyle{plain}

\begin{thebibliography}{M}
%
\bibitem{ZZ1}
I. Ziemann, V. Ziemann, {\em Noninvasively improving the orbit-response matrix
  while continuously correcting the orbit,} Physical Review Accelerators and
Beams 24 (2021) 072804
\bibitem{MIZI}
  M. Minty, F. Zimmermann, {\em Measurement and Control of Charged Particle Beams,}
  Springer, Heidelberg, 2003.
\bibitem{HUANG}
  X. Huang, {\em Beam-based correction and optimization for accelerators,}
  CRC press, Boca Raton, 2020.
\bibitem{CORBETT}
  J. Corbett et al., {\em A Fast Model Calibration Procedure for Storage Rings,}
  Proceedings of the Particle Accelerator Conference PAC93, Washington, 1993, p. 108.
\bibitem{SAFRANEK}
  J. Safranek, {\em Experimental determination of storage ring optics using orbit
    response measurements,} Nuclear Instruments and Methods A 388 (1997) 27.
\bibitem{SYSINF}
  G. Goodwin, R. Payne, {\em Dynamic System Identification,} Academic Press, London, 1977.
\bibitem{LJUNG}
  L. Ljung, {\em System Identification; theory for the user, 2nd ed.}, Prentice Hall,
  New Jersey, 1999.
\bibitem{FYF}
  V. Ziemann, {\em Regression Models and Hypothesis Testing.} In: Physics and Finance.
  Undergraduate Lecture Notes in Physics. Springer, Cham.
  \url{https://doi.org/10.1007/978-3-030-63643-2_7}.
\bibitem{SHEMO}
  W. Press et al., {\em Numerical Recipes, 2nd ed.,} Cambridge University Press, Cambridge, 1992.
\bibitem{SYSINF2}
  Section 7.3 in~\cite{SYSINF}.

%
\end{thebibliography}

\end{document}